\documentclass[a4paper,11pt]{article}
\usepackage{amsfonts}
\usepackage{amsmath, amsfonts, amsthm, amssymb, amscd, mathrsfs}
\usepackage[mathcal]{euscript}
\usepackage[toc,page]{appendix}
\newcounter{mnotecount}[section]

\renewcommand{\themnotecount}{\thesection.\arabic{mnotecount}}

\newcommand{\mnote}[1]%{}%
{\protect{\stepcounter{mnotecount}}$^{\mbox{\footnotesize
$%\!\!\!\!\!\!\,
\bullet$\themnotecount}}$ \marginpar{%\color{red}%
\raggedright\tiny\em
$\!\!\!\!\!\!\,\bullet$\themnotecount: #1} }

\usepackage{ae}

\usepackage[ansinew]{inputenc}

\usepackage{lscape}

\usepackage{amssymb,amsmath,amsthm}
\usepackage{systeme}

\usepackage{epsf}

\usepackage{psfrag}

\usepackage{mathrsfs}

\usepackage{xcolor}

\theoremstyle{definition}

\newtheorem{thm}{Theorem}

\newtheorem{assumption}{Assumption}

\newtheorem{Remark}{Remark}

\title{Energy bounds and ergoregion instability in Einstein-Maxwell-Scalar field models}

\author{Filipe C. Mena$^{(1,2)}$\footnote{email: filipecmena@tecnico.ulisboa.pt} and Jo\~ao  M. Oliveira$^{(2)}$\\\\
{\small $^{(1)}$ Centro de An\'alise Matem\'atica, Geometria e Sistemas Din\^amicos,}
\\
{\small Instituto Superior T\'ecnico, Universidade de Lisboa, Av. Rovisco Pais 1, 1049-001 Lisboa, Portugal}
\\
{\small $^{(2)}$ Centro de Matem\'atica, Universidade do Minho, 4710-057 Braga, Portugal}
}

\begin{document}

\maketitle

\begin{abstract}
We investigate energy bounds and the stability of stationary asymptotically flat spacetimes with an ergoregion and no future horizon in the context of Einstein-Maxwell-Scalar field models which naturally arise in Kaluza-Klein and String theories. In order to do that we consider scalar field perturbations non-minimally coupled to a background electromagnetic field. We show that there is compactly supported initial data such that the initial energy flux across the initial Cauchy hypersurface is negative and that the energy flux decreases with time. Then, considering arguments due to J. Friedman and G. Moschidis, this indicates that the energy flux diverges for those solutions and such spacetimes with an ergoregion are unstable.
\end{abstract}
\newpage
%%%%%%%%%%%%%%%%%%%%%%%%%%%%%%%%%%%%%%%%%%
\section{Introduction}

 Ergoregions play an important role in astrophysics as they can account for observable differences between black holes and horizonless ultra-compact objects \cite{Cardoso-2008} and they may impact on the formation of powerful relativistic jets \cite{Shapiro-2020, Shapiro-2019}. 

The study of the stability of spacetimes with an ergosphere and without horizons is therefore interesting as they include, for example, models of neutron stars and boson stars. In this context, an important result was achieved by J. Friedman \cite{Friedman} who showed that the ergoregion of horizonless stationary asymptotically flat spacetimes is unstable under scalar field or electromagnetic  linear perturbations of high enough frequency, 
in the sense that the energy of the solutions to the respective wave equation grows unbounded in time. This result has been rigorously proved in a notable paper of G. Moschidis \cite{Moschidis}.

From the numerical point of view, the ergosphere stability of stationary rotating horizonless objects was then studied for neutron stars \cite{Comins, Yoshida, Kokkotas}, gravastars \cite{Rezzolla}, boson stars \cite{Cardoso-2022}, Kerr-like horizonless objects \cite{Maggio2, Maggio1} and hydrodynamic vortex \cite{Cardoso-2014}. In those studies, particular attention has been paid to the stability time-scales of ergoregions  specially in connection with superradiance instabilities (see \cite{Brito} for a review and further references).

The above results were obtained in the context of General Relativity. Given the recent interest in black holes and other compact objects with ergoregions in more general theories of gravity, in this paper we investigate the stability of ergoregions in theories with a non-minimal coupling between a scalar field and a Maxwell field, i.e. in the so-called Einstein-Maxwell-Scalar (EMS) theories. A physical motivation to study such theories is that they arise naturally in the context of Kaluza-Klein models \cite{Appelquist:1987nr} and String Theory \cite{VanNieuwenhuizen:1981ae}. In those contexts, the coupling function is typically an exponential function \cite{GibMae.1988} but more general classes of coupling functions have also been considered, for example, in cosmology \cite{Martin:2007ue, Maleknejad:2012fw}.
Some solutions found in EMS models representing horizonless physical systems include for instance some dynamical boson stars \cite{Boson}, stationary rotating G\"odel-type models \cite{Ishihara} as well as solitons \cite{Herdeiro:2019iwl, Herdeiro:2019oqp,Herdeiro:2020iyi}.

The EMS theories in general give rise to a coupled system of non-linear wave equations for the scalar field and the electromagnetic field. While General Relativity leads to decoupled linear homogeneous wave equations, the EMS theories include non-linear terms resulting from products of the coupling function and first derivatives of the fields. So, the study of the stability of general solutions to the full system is highly non-trivial and has not been attempted yet. 

Then, as a first step, in this paper we will consider that the scalar field acts as a perturbation on an electromagnetic spacetime background which is stationary and asymptotically flat. By retaining the lowest order terms in the perturbation, we will consider a single wave equation of Klein-Gordon type with variable coefficients. So it is not straightforward to see {\em a priori} if the methods of Friedman and Moschidis work in this case since, as part of their instability proof, the sign of an energy functional depending on the stress-energy tensor has to be controlled and, in our case, that functional includes extra terms from the EMS theory.

Interestingly, we find that if the function coupling the scalar field and the electromagnetic field is appropriately constrained, one can indeed control the coupling terms which appear in the new energy functionals and find initial data such that the energy flux through spacelike Cauchy surfaces does not decay to zero as time increases. Then, relying on Friedman's arguments and results of Moschidis, which in part apply also to our Klein-Gordon equation, our result suggests that the energy flux diverges for those solutions and the considered spacetimes with an ergoregion are unstable.

The plan of the paper is as follows: In Section \ref{EMS} we describe the Einstein-Maxwell-Scalar models. In Section \ref{set-up} we state our assumptions and the main result of the paper. The proof of the main result is contained in Section \ref{proof}, while in Section \ref{finite-energy} we comment on the instability for our setting. Finally, we make concluding remarks and mention open problems in Section \ref{conclusion}.

We use units such that $8\pi G=c=1$ and consider greek indices $\alpha, \beta,... \mu, \nu,...=0,1,2,3$. 
%%%%%%%%%%%%%%%%%%%%%%%%%%%%%%%%%%%%%%%%%%
\section{Einstein-Maxwell-Scalar (EMS) models}
\label{EMS}

The Einstein-Maxwell-Scalar models have a stress-energy tensor of the form
\begin{equation}
\label{stress-energy}
\mathcal{T}_{\mu\nu}=f(\phi)(F_{\mu\alpha} F_{\nu}^{~\alpha}-\frac{1}{4}g_{\mu\nu}F_{\alpha\beta}F^{\alpha\beta})+\partial_\mu\phi\partial_\nu\phi-\frac{1}{2}g_{\mu\nu}(\partial\phi)^2 ,
\end{equation}
where $(\partial \phi)^2 := g_{\mu\nu}\partial^\mu \phi \partial^\nu \phi$ and
$$F_{\mu\nu} = (dA)_{\mu\nu}$$
 is the Faraday tensor of electromagnetism,
$A^\alpha$ the usual vector potential in the Lorentz gauge, $\phi$ a real scalar field and $f$ a smooth (coupling) function that we choose to be positive. 
Furthermore, we assume that $f$ has the form
\begin{equation}
\label{ansatz-f}
f(\phi)=1+\lambda h(\phi)>0,
\end{equation}
where $|\lambda|<1$ and $h$ is a smooth function such that  
\begin{equation}
\label{condition}
 \lim_{\substack{\phi\to 0}} h(\phi)=0
\end{equation}
so that we recover the Einstein-Maxwell case in the absence of the scalar field.
We note that conditions \eqref{ansatz-f}-\eqref{condition} include the main cases studied so far in the literature \cite{GibMae.1988,Herdeiro:2019iwl,Martin:2007ue, Maleknejad:2012fw}.

In what follows it is useful to consider three parts of the stress-energy tensor \eqref{stress-energy} as:
\begin{equation}
\aligned
\label{Tdecomp}
(\mathcal{T}_{\phi})_{\mu\nu} &= \partial_\mu\phi\partial_\nu\phi-\frac{1}{2}g_{\mu\nu}(\partial\phi)^2  \ ,\\
(\mathcal{T}_{\mathrm{EM}})_{\mu\nu} &= F_{\mu\alpha} F_{\nu}^{~\alpha}-\frac{1}{4}g_{\mu\nu}F_{\alpha\beta}F^{\alpha\beta} \ ,\\
(\mathcal{T}_{\mathrm{EMS}})_{\mu\nu} &=(\mathcal{T}_{\mathrm{EM}})_{\mu\nu} \lambda h \ , 
\endaligned
\end{equation}
which we call, respectively, the scalar, electromagnetic and Einstein-Maxwell-Scalar parts. 
 
In EMS theories with stress-energy tensor \eqref{stress-energy}, while $f$ is given a priori, $\phi$ and $A_{\mu}$ must satisfy the following non-linearly coupled system of five second order partial differential equations \cite{Herdeiro:2018wub}:
\begin{eqnarray}
\label{wave-eq}
\Box_g \phi&=&\displaystyle{\frac{1}{4}f'(\phi)\left(g_{\alpha\mu}g_{\beta\nu}\nabla^{[\mu} A^{\nu]}\nabla^{[\alpha} A^{\beta]}\right)}, \\
\label{wave-A}
\Box_g A^{\alpha}&=&-\displaystyle{\frac{f'(\phi)}{f(\phi)}\nabla_\mu\phi\nabla^{[\alpha } A^{\mu ]}} \ .
\end{eqnarray}
where $\Box_g:= \nabla_\mu\nabla^\mu$ is the wave operator with respect to the metric $g$, the prime denotes differentiation with respect to $\phi$ and the squared brackets denote index anti-symmetrization.

While the previous wave equations generalise the homogeneous equation $\Box_g \phi=0$ often studied in General Relativity, the appearance of the inhomogeneous and non-linear term brings new mathematical challenges. Assuming that the coefficient functions of the system are regular then, by standard results about semi-linear hyperbolic PDEs (see e.g. \cite{Choquet}), the system \eqref{wave-eq}-\eqref{wave-A} has a locally well-posed Cauchy problem. 
It is also known that the system admits explicit solutions compatible with asymptotic flatness \cite{Herdeiro:2018wub, Herdeiro:2019iwl, Herdeiro:2019oqp, Herdeiro:2020iyi}.

Here we will consider a model where the electromagnetic field is time-independent and is part of the spacetime background where $\phi$ acts as a perturbation. So we will consider that $A^\alpha$ is given {\em a priori} and   \eqref{wave-A} will not be used. This is will be more clearly stated in the assumptions of the next section.

%%%%%%%%%%%%%%%%%%%%%%%%%%%%%%%%%%
\section{Geometric setup and main result}
\label{set-up}
%%%%%%%%%%%%%%%%%%%%%%%%%%%%%%%%%%%
We start by stating our assumptions which are similar to Moschidis's \cite{Moschidis} except for the new spacetime model (item 5 below) and the asymptotic regularity (item 6, which was also considered by Friedman \cite{Friedman}):
\begin{assumption} 
\label{assump1}
We consider a smooth globally hyperbolic 4-dimensional spacetime $(M,g)$ such that:
\begin{enumerate}

\item {\bf (Stationarity)} There exists a Killing vector field $T$ with complete orbits and a smooth Cauchy hypersurface $\Sigma\subset M$ such that $T |_{\Sigma}$ is transversal to $\Sigma$ and is timelike outside a compact subset of $\Sigma$. 

\item {\bf (Asymptotic flatness)} $(M,g)$ is asymptotically flat having the following metric form
\begin{equation}
\label{metric} 
g= -(1+O(r^{-1}))dt^2+(1+O(r^{-1}))dr^2 +r^2 (g_{{\mathbb S}^{2}}+O^{{\mathbb S}^{2}}(r^{-1}))+O(1)du d\Omega,
\end{equation}
where $g_{{\mathbb S}^{2}}$ denotes the metric on the $2$-sphere, $d\Omega$ denotes an angle differential and $O^{{\mathbb S}^{2}}(r^{-1})$ a tensor on ${\mathbb S}^{2}$ with a $O(r^{-1})$ decay. 
\item {\bf (Ergoregion)} $(M,g)$ contains an ergoregion 
$${\mathscr E}:= \overline {\{ g(T,T)>0  \}}\ne \emptyset$$
so that $T$ is timelike on the complement of ${\mathscr E}$ and the boundary $\partial {\mathscr E}$ is a smooth hypersurface.
\item {\bf (No horizons)} $(M,g)$ does not contain future event horizons. 
\item {\bf (Probe scalar field)} There is a smooth scalar field perturbation $\phi$, small enough, on a given spacetime background with a time-independent electromagnetic field with vector potential $A^\mu$. At lowest order, $\phi$ satisfies a (linear) Klein-Gordon type equation
\begin{equation}
\label{wave-eq-2}
\Box_g \phi- \lambda H\phi=F,
\end{equation}
where $|\lambda|<1$ and $H$ and $F$ are time-independent functions with sufficiently fast decay at infinity. 

\item{\bf (Asymptotic regularity)} We assume the following expansions in inverse powers of $r$ in a neighbourhood of future null infinity $\mathscr{I}^+$:
\begin{align}
\label{expansion-phi}
\phi= \sum_{n=1}^{\infty} \frac{\phi_n(u,\theta,\varphi)}{r^n},~~~~~~~A^\alpha= \sum_{n=1}^{\infty} \frac{A^\alpha_n(\theta,\varphi)}{r^n}.
\end{align}

\item {\bf (Unique continuation across $\partial {\mathscr E}$)} There exists an open neighborhood $\cal U$ of a point $p \in\partial {\mathscr E}$ such that any solution to \eqref{wave-eq-2} with $\phi\equiv 0$ in $M\setminus {\mathscr E}$ also satisfies $\phi\equiv 0$ in ${\mathscr E} \cap \cal U$.

\end{enumerate}
\end{assumption}
%%%%%%%%%%%%
\begin{Remark}
\label{remark1}
Condition 7 of Assumption \ref{assump1} holds under the existence of an axisymmetric Killing vector field $\Phi$ such that the span of $\Phi$ and $T$ is timelike on $\partial {\mathscr E}$ \cite{Moschidis}.
This unique continuation criterium was also assumed by Moschidis \cite{Moschidis} and allows the use of previous results of Tataru \cite{Tataru} regarding the continuation of solutions to wave equations. Furthermore, Condition 7 is automatically satisfied if the spacetime is analytic, as in the case of Friedman \cite{Friedman}.
\end{Remark}

Now, considering the Killing vector field $T$ and its properties defined above we have that
near infinity $T$ is timelike. We then assume without loss of generality the asymptotic norm 
$g(T,T)=-1.$
Take an initial Cauchy surface  $\Sigma_0$ and a family of spacelike hypersurfaces $\Sigma_u$ indexed by the scalar $u$ such that 
$$g(T, \nabla u) = 1,$$
so $\Sigma_u$ is the image of $\Sigma$ under the flow of $T$ for each $u$.
Defining
\begin{equation}
\label{mu-eq}
\mu=\sqrt{-(\nabla u)^2},
\end{equation}
then the (future directed) unit normal to each $\Sigma_u$ is
$$n_{\Sigma_u} = -\frac{1}{\mu}\nabla u,$$
which has norm $g(n_{\Sigma_u}, n_{\Sigma_u}) = -1$. When it is clear from the context we will write $n_{\Sigma_u}$ simply as $n$. 

Taking into account the previous notation, we now present the main result of the paper:
\begin{thm}
\label{main-thr}
Consider a spacetime $(M,g)$ satisfying Assumption \ref{assump1}. Then:
\begin{enumerate}
\item The $T$-energy flux 
\begin{equation}
\label{conserved-current}
J_\mu^T(\phi, A^\alpha):= \mathcal{T}_{\mu\nu}T^\nu
\end{equation}
satisfies 
\begin{equation}
\label{part1-eq}
\int_{\Sigma_u} J_\mu^T(\phi, A^\alpha) n^\mu_{\Sigma_u} \le \int_{\Sigma_0} J^T_\mu(\phi, A^\alpha)  n^\mu_{\Sigma_0}
\end{equation}
where $\Sigma_0$ is an initial Cauchy surface and $\Sigma_u$ the image of $\Sigma$ under the $T$-flow for time $u$. 
\item Given smooth $A^\alpha$ and $f(\phi)$, there is compactly supported initial data on $\Sigma_0$ such that the initial energy is negative and
\begin{equation}
\label{negativity}
\int_{\Sigma_u \cap {\mathscr E}} J_\mu^T(\phi, A^\alpha) n^\mu_{\Sigma_u} \le \int_{\Sigma_0} J_\mu^T(\phi, A^\alpha) n^\mu_{\Sigma_0}<0.
\end{equation}
\end{enumerate}
\end{thm}
The proof of parts 1 and 2 of the theorem are detailed in sections \ref{part1} and \ref{part2}, respectively. Using results of Friedman \cite{Friedman} and Moschidis \cite{Moschidis}, see Section \ref{finite-energy}, this result suggests that the energy flux diverges at $\mathscr{I}^+$ i.e.
\begin{equation}
\label{instability}
\int_{\mathscr{I}^+} J_\mu^T(\phi,A^\alpha) n^\mu_{\mathscr{I}^+} =+\infty
\end{equation}
so that the considered spacetimes with ergoregion are unstable.
%%%%%%%%%%%%%%%%%%%%%%%%%%%%%%%%%%%%%%%%%%%%%%%%%%%%%
\section{Energy integrals}
\label{sec-energy}

In this section we construct energy integrals that will be useful later. 
Consider a hypersurface $\Sigma_u$ with future directed unit normal $n_{\Sigma_u}$. We recall that the electric and magnetic fields with respect to $n_{\Sigma_u}$ are given by
$$E = F(n_{\Sigma_u}),~~~~~B = (\star F)(n_{\Sigma_u}),$$
where $\star$ denotes the Hodge operator. The Faraday tensor can thus be written as
\begin{equation}
\label{faraday-t}
F^{\mu\nu} = 2n^{[\mu}E^{\nu]}+\varepsilon^{\mu\nu\alpha\beta}B_\alpha n_\beta
\end{equation}
so that
$
F^2:= F_{\alpha\beta}F^{\alpha\beta}=2(B^2-E^2).
$
It also useful to define
$$\tilde E = F(T),~~~~~\tilde B = (\star F)(T)$$
and, using the identities, $\nabla_{[\mu} F_{\alpha\beta]}=0$, we obtain
$\mathcal{L}_T F = d\tilde{E}$
and
$ \mathcal{L}_T(\star F) = d\tilde{B}$. 
Furthermore, to ease the notation for the directional derivative along a vector field $X$ we use 
$$
X\phi:= X^\mu \nabla_\mu \phi.
$$ 
Recalling the (conserved) current \eqref{conserved-current} and using \eqref{stress-energy}, we find
\begin{equation}
\label{current2}
(J^T)^\mu(\phi, A^\alpha)=f(\phi)\left[F^{\mu\nu} \tilde E_\nu-\frac{1}{4} F^2 T^\mu\right]-\frac{1}{2}(\partial\phi)^2 T^\mu+ (T\phi) \nabla^\mu\phi.
\end{equation}
In turn, we define the energy flux on each surface $\Sigma_u$ as
\begin{equation}
\label{energy-func}
{\cal E}_u:=\int_{\Sigma_u} J^T_\mu(\phi,A^\alpha) n^\mu_{\Sigma_u}
\end{equation}
giving
\begin{align}
\label{EIntS}
\mathcal{E}_u = 
\int_{\Sigma_u} \left[\frac{f(\phi)}{2}(E^\mu \tilde{E}_\mu + B^\mu  \tilde{B}_\mu)-\frac{1}{2} (\partial\phi)^2 T_\mu n^\mu 
+ (T\phi)(n \phi) \right] dS,
\end{align}
which, in the absence of a scalar field, reduces to the expression obtained by Friedman \cite{Friedman}.
However, as opposed to the Einstein-Maxwell model of \cite{Friedman}, in our case it is not immediately clear that time independent configurations of the fields imply that the energy of the system is zero for asymptotically flat spacetimes.

In order to explore this further we write the vector fields $E$ and $B$ with respect to the vector potential $A$ and another potential $\hat{A}$ as
\begin{align}
E  = (\star d A)(n_{\Sigma_u}), ~~~~~
B  = (\star d\hat A) (n_{\Sigma_u})\ .
\end{align}
With these expressions we can integrate by parts and re-write the integral \eqref{EIntS} as:
\begin{align}
\mathcal{E}_u =& \frac{1}{2}\int_{\Sigma_u} \left[(T\phi)(n\phi)-\phi n^\mu T\nabla_\mu\phi+\varepsilon^{\mu\nu\alpha\beta}n_\nu \frac{f(\phi)}{2}\left( A_\beta \mathcal{L}_T F_{\mu\alpha}+\hat{A}_\beta \mathcal{L}_T (\star {F})_{\mu\alpha} \right)\right] dS \nonumber \\
&+ \int_{\partial \Sigma_u}\left[\phi (n \phi) T^\mu - f(\phi)\varepsilon^{\mu\nu\alpha\beta}n_\nu\left(A_\beta \tilde{E}_\alpha+\hat{A}_\beta \tilde{B}_\alpha \right)\right]dS_\mu + \mathcal{E}_{\mathrm{EMS}} \ ,\label{TEnergy}
\end{align}
with
\begin{equation}\label{EEMS}
\mathcal{E}_{\mathrm{EMS}} = \int_{\Sigma_u} f'(\phi)\left[\frac{1}{8}\phi   F^2 g(T, n)  + \varepsilon^{\mu\nu\alpha\beta}n_\nu \nabla_\mu \phi \left(A_\beta\tilde{E}_\alpha+ \hat{A}_{\beta}\tilde{B}_\alpha \right)\right]dS,
\end{equation}
where, as in \eqref{EIntS}, we used $n\equiv n_{\Sigma_u}$ to simplify the notation. 

So we see that, for asymptotically flat and time independent configurations satisfying \eqref{expansion-phi}, all the contributions except for the pure EMS contribution will disappear. This means that, unlike the case with $\Box_g \phi=0$, in our case there can be a time independent configuration with non-zero energy $\mathcal{E}^s$, to which the system can settle down after radiating. This energy state can have either negative or positive energy and this will be relevant during the proof of part 2 of Theorem \ref{main-thr}.
%
%%%%%%%%%%%%%%%%%%%%%%%%%%%%%%%%%%%%%%%%%%%%%%%%%
\section{Proof of Theorem 1}
\label{proof}
\subsection{Proof of part 1}
\label{part1}

Take a spacetime region $\cal{R}$ bounded by two spacelike Cauchy surfaces $\Sigma_0$ and $\Sigma_u$ and the causal boundary surface $\Sigma_c$ connecting them. The normal vector to this connecting surface is the following null vector
\begin{equation}
N = \partial_t + \partial_r = \partial_u + 2 \partial_r.
\end{equation}
Because $J_\mu^T(\phi,A^\alpha)$ is conserved on the whole region, we know that
\begin{align}
\int_\mathcal{R} \nabla^\mu J^T_\mu(\phi,A^\mu) dV 
&= \int_{\Sigma_u} J^T_\mu(\phi,A^\mu)  n_{\Sigma_u}^\mu  - \int_{\Sigma_0} J^T_\mu(\phi,A^\mu)  n_{\Sigma_u}^\mu  + \int_{\Sigma_c} J^T_\mu(\phi,A^\mu) N^\mu  \nonumber\\ &= \mathcal{E}_u - \mathcal{E}_0 + \mathcal{E}_c = 0 \ ,
\end{align}
so that we integrate in $\mathbb S^2\times [0,u] $, i.e. over the 2-spheres for each $u$.
A calculation from \eqref{current2} gives
\begin{align}
J^T_\mu (\phi, A^\alpha) N^\mu&=(T\phi)(T\phi+\nabla_r\phi)+f(\phi) F(\tilde E)_\mu N^\mu  \ .
\end{align}
From the assumption \eqref{expansion-phi} regarding the regularity of $\phi$ and $A^\alpha$ near future null infinity, we have
\begin{align}
\phi(u,r,\theta,\varphi)&= \frac{1}{r}\phi^0(u,\theta,\varphi) + \mathcal{O}(r^{-2}) \ ,	 \\
A_\mu(r,\theta,\varphi) &= \frac{1}{r}A^0_\mu(\theta,\varphi) + \mathcal{O}(r^{-2}) \ ,
\end{align}
then around $\mathscr{I}^+$
\begin{equation}
f(\phi) F(\tilde E)_\mu N^\mu \sim f(\phi)\mathcal{O}(r^{-3}) \ .
\end{equation}
We can now calculate the energy difference between the surfaces $\Sigma_0$ and $\Sigma_u$ as
\begin{align}
{\cal E}_u-{\cal E}_0=-\lim_{r\to\infty} \int_0^u \int_{\mathbb S^2}J^T_\mu (\phi,A^\mu)N^\mu r^2 d\Omega_{\mathbb S^2} du 
			      =-\lim_{r\to\infty} \int_0^u \int_{\mathbb S^2}[(T\phi)^2 + f(\phi) \mathcal{O}(r^{-3})]r^2 d\Omega_{\mathbb S^2}  du
			      \nonumber
\end{align}
which is negative. This means that the system radiates energy away between $\Sigma_0$ and $\Sigma_u$. Then, we can write that the $T$-energy flux satisfies \eqref{part1-eq}
 which concludes the proof of part 1.
%
%%%%%%%%%%%%%%%%%%%%%%%%%%%%%%%%%%%%%%%%%%
\subsection{Proof of part 2}
%%%%%%%%%%%%%%%%%%%%%%%%%%%%%%%%%%%%%%%%%%
\label{part2}

Let $n:=n_{\Sigma_u}$ to simplify the notation in this section. Following Friedman \cite{Friedman}, we decompose $T$ as
\begin{equation}
T = \frac{1}{\mu}(n +\alpha k) \ ,
\end{equation}
where $\mu$ was defined in \eqref{mu-eq} and $k$ is a vector field orthogonal to $n$ that respects
$$
g(k, n) = 0~~~~ {\text {and}}~~~~ g( k, k) = 1.
$$
In turn 
$\alpha>0$
is an important parameter that defines the character of $T$ in the sense that if $\alpha>1$, $\alpha<1$ or $\alpha=1$, respectively, then $T$ is spacelike, timelike or null. In particular, $\alpha$ signals the presence of the ergoregion where $\alpha>1$. 

We consider an useful projection operator orthogonal to $n$ and $k$ as
\begin{equation}
j^{~\mu}_\nu := \delta^{~\mu}_\nu + 4 n^{[\mu} k^{\alpha]} n_{[\nu} k_{\alpha]} \ ,
\end{equation}
where $\delta^{~\mu}_\nu$ is the usual delta Kronecker symbol.

We can now write the total energy \eqref{energy-func} as
\begin{equation}
\mathcal{E}_u = \int_{\Sigma_u} \frac{1}{\mu}\mathcal{T} (n+\alpha k,n) \,dS\ ,
\end{equation}
where we recall that $\mathcal{T}$ is the energy-momentum tensor \eqref{stress-energy} and, using \eqref{faraday-t}, we can write the integrand as
\begin{align}
\label{Integrand}
\mathcal{T} (n+&\alpha k, n) =\nonumber\\
=& f(\phi)\left[F^{\mu\alpha}F_{\nu\alpha}(n^\nu+\alpha k^\nu) n_\mu + \frac{1}{4}F^2\right]+(n\phi)^2 \nonumber
+ \alpha(n\phi)(k\phi)+\frac{1}{2}(\partial\phi)^2 \nonumber \\
=& f(\phi)\left[\frac{1}{2}\left(E^2+B^2\right)+\alpha\varepsilon^{\alpha\beta}E_\alpha B_\beta\right]+ \frac{1}{2}(n\phi)^2 +\frac{1}{2}(k\phi)^2+\alpha(n\phi)(k\phi)+\frac{1}{2}j^{\alpha\beta}\partial_\alpha\phi\partial_\beta\phi ,
\end{align}
where $\varepsilon^{\alpha\beta} = k_\mu n_\nu \varepsilon^{\mu\nu\alpha\beta}$.

As an aside, we note that since 
$|\varepsilon^{\alpha\beta}E_\alpha B_\beta|\leq \sqrt{g(E, E)}\sqrt{g(B, B)},$
then the dominant energy condition $\mathcal{T}(T, n) > 0$ is verified whenever $\alpha\leq 1$, i.e. when there is no ergoregion. As we shall see, when we have an ergoregion we can find initial conditions for which $\mathcal{T}(T, n) < 0$.

We now follow part of Friedman's strategy and generalise it to our case in order to find appropriate initial conditions that prove \eqref{negativity}.

Consider an open set $\Omega\subset {\mathscr E}$ set and choose a chart $(t,x,y,z)$ such that curves of constant $t, y, z$ have tangent $k$ with
$\partial_x := k^\mu\nabla_\mu.$
 For a small $\epsilon$, we have $\alpha=1+\epsilon>0$ on $\Omega$. Consider now a ball $B_R(p)\subset \Omega$, with radius $R$ centered at the point $p\in\Omega$. Let $\rho\in C^\infty(\Omega)$ be a function such that $\rho=1$ on $B_R(p)$ and $\rho=0$ outside a compact subset of $\Omega$. This function and its derivatives are assumed to be bounded initially by a constant $K>0$ as
\begin{equation}\label{BoundK}
\sup_\Omega|\rho| + \sup_\Omega|k\rho| + \sup_\Omega\sqrt{j^{\mu\nu}\nabla_\mu\rho\nabla_\nu\rho} < K \ .
\end{equation}
We also consider that the electromagnetic term of \eqref{Integrand} is bounded by some $L>0$ as
\begin{equation}
\sup_\Omega|E^2+B^2+2\alpha\varepsilon^{\alpha\beta}E_\alpha B_\beta| < L \ .
\end{equation}
Taking into account these conditions, following \cite{Friedman}, we now consider the initial data for the scalar field
\begin{equation}
\aligned
&\phi_m = \rho\sin(mx) \label{Init1}\\
&(n+k)\phi_m = 0.
\endaligned
\end{equation}
We will show that these conditions give $\mathcal{E}_\phi<0$ on $\Sigma_u$, for large enough $m$ (representing 
a perturbation with high enough frequency), where
\begin{equation}
\mathcal{E}_\phi := \mathcal{E}_u - \mathcal{E}_{\mathrm{EM}} = \int_{\Sigma_u}  \mathcal{T}(T,n)dS-\int_{\Sigma_u}\mathcal{T}_{\mathrm{EM}}(T,n)dS
\end{equation}
is the energy contribution of the scalar field perturbation (recall our definitions \eqref{Tdecomp}).
We note that since we consider the background fixed, the electromagnetic energy will not have an impact in the following argument.

Using the above initial conditions, the expression \eqref{Integrand} becomes
\begin{align}
\mathcal{T}(T,n) = \mathcal{T}_{\mathrm{EM}}(T,n) + \mathcal{T}_{\mathrm{EMS}}(T,n)+ (1-\alpha)(k\phi_m) + \frac{1}{2}j^{\mu\nu}\nabla_\mu\phi_m\nabla_\nu\phi_m\ .
\end{align}
Recall that $\mu=\sqrt{-(\nabla u)^2}$ which is bounded on $\Omega$. Denoting the initial bounds of the function $\mu$ on $\Omega$ as
$
0<\mu_0\leq\mu\leq\mu_1 \ ,
$
we find
\begin{align}
\mathcal{E}_\phi &= \int_{\Sigma_u} \frac{1}{\mu}( \mathcal{T}-\mathcal{T}_{\mathrm{EM}})(T,n)\, dS \nonumber \\
&\leq\int_\Omega\frac{1}{\mu}\mathcal{T}_{\mathrm{EMS}}(T,n)\, dS-\frac{\epsilon}{\mu_1}\int_{B_R(p)}(k\phi_m)^2 dS +\frac{1}{\mu_0}\int_{\Omega}j^{\mu\nu}\nabla_\mu\phi_m\nabla_\nu\phi_m dS \nonumber\\
&\leq\frac{1}{2}\int_\Omega \frac{\lambda h(\phi_m)}{\mu}(E^2+B^2+2\alpha \varepsilon^{\alpha\beta}E_\alpha B_\beta)dS-\frac{\epsilon}{\mu_1}\int_{B_R(p)}(k\phi_m)^2 dS +\frac{1}{\mu_0}K^2 \text{Vol}(\Omega)  \ .
\end{align}
The first integral in the previous expression corresponds to the total EMS contribution to the energy and includes the extra term \eqref{EEMS}. In this form we can see that this extra energy only depends on powers of $\phi_m$ and not on its derivatives. We now assume that $f(\phi_m)$ is bounded on $\Omega$ as
$
\sup_\Omega |\lambda h(\phi_m)|\leq C \ ,
$
for some $C>0$. Then
\begin{align}
\int_\Omega \frac{\lambda h(\phi_m)}{\mu}(E^2+B^2+2\alpha \varepsilon^{\alpha\beta}E_\alpha B_\beta)dS 
&\leq \frac{C}{\mu_1}L\int_\Omega dS = \frac{C}{\mu_1}L \text{Vol}(\Omega) \ . \label{Eemsb}
\end{align}
So we obtain
\begin{align}
\mathcal{E}_\phi 
&\leq  \frac{C}{2\mu_1}L \text{Vol}(\Omega)-m^2\frac{\epsilon}{2\mu_1} \text{Vol}(B_R(p)) +\frac{1}{\mu_0}K^2 \text{Vol}(\Omega) \ , \label{Ephibound}
\end{align}
where we used the fact that, for $m\rightarrow \infty$ we have
$
\int_{B_R(p)}\cos^2(mx) dS \rightarrow \frac{1}{2} \text{Vol}(B_R(p)).
$
We can always find initial conditions with sufficiently large $m$ such that the middle term in \eqref{Ephibound} is the largest in magnitude (implying $\mathcal{E}_\phi<0$) and large enough so that the total energy $\mathcal{E}_u$ is negative. 
Then, taking into account the previous calculations for the initial data considered, along with the conclusion \eqref{part1-eq} of part $1$, we have \eqref{negativity}
 showing part $2$ of the theorem.

\begin{Remark}
At the end of Section \ref{sec-energy} we mentioned that for our EMS model, we can have a finite and time-independent negative energy state $\mathcal{E}^s<0$ due to the extra term \eqref{EEMS}. This means that, if the initial conditions correspond to a state $\mathcal E$ such that $\mathcal E^s<\mathcal E<0$, the system might settle down to the time-independent state $\mathcal E^s$. However, we can always choose $m$ large enough such that $\mathcal E<\mathcal E^s<0$.
This is because the term \eqref{EEMS} 
 is contained in the interaction energy given by equation \eqref{Eemsb} which is always bounded. When this happens, we have a radiating state that can never settle into a time-independent state. 
\end{Remark}
%
%%%%%%%%%%%%%%%%%%%%%%%%%%%%%%%%%%%%%%%%%%%%%%%
\subsection{Cases with $f=f(\phi,(\partial \phi)^2)$}
As an aside, in this subsection, we make a comment regarding a possible generalisation of the result of Section \ref{part2} by considering functions more general than \eqref{ansatz-f}.

In fact one might try to circumvent the result of the previous section by considering a coupling function $f$ that also depends on the derivatives of the scalar field i.e.
\begin{equation}
f(\phi,(\partial \phi)^2)=1+\lambda h(\phi,(\partial \phi)^2)>0,
\end{equation}
where $|\lambda | <1$ as before and $h$ is a smooth function such that  
\begin{equation}
 \lim_{\substack{\phi\to 0 \\ \partial \phi\to 0}} h(\phi,(\partial \phi)^2)=0,
\end{equation}
mimicking conditions \eqref{ansatz-f}-\eqref{condition} and assuring a well defined General Relativity limit as $|\lambda|\to 0$.

Just like the coupling $f$ depending on $\phi$ can give rise to scalarised solutions originating from a tachyonic instability, a coupling depending on derivatives of $\phi$ can create a so-called \emph{ghostlike} instability (see e.g. \cite{Ramazanoglu:2019gbz}, section 4, for details). So, from the physical point of view, this is an interesting case. 

In fact, after repeating the steps of the last section, we can see that the derivative terms give new terms that depend on $m^2$ and one may ask if those terms can counteract the negative term in \eqref{Ephibound}. However, this is not the case. To see why, we start by expanding $f$ as
\begin{equation}
f(\phi,(\partial \phi)^2) =1+\lambda \frac{\partial h}{\partial\phi}(0,0)\phi +\lambda \frac{\partial h}{\partial((\partial \phi)^2)}(0,0)(\partial \phi)^2 + \mathcal{O}(\phi^2, (\partial \phi)^4) \ .
\end{equation}
Now, assuming that the derivative terms are regular at $(0,0)$, we can write $(\partial \phi)^2$ as
\begin{align}
(\partial \phi)^2 = \nabla_\mu\phi\nabla^\mu\phi =  (j^{~\mu}_\nu - 4 n^{[\mu} k^{\alpha]} n_{[\nu} k_{\alpha]})\nabla_\mu\phi \nabla^\nu\phi 
= j^{~\mu}_\nu\nabla_\mu\phi \nabla^\nu\phi-(n\phi)^2 + (k\phi)^2 \ .
\end{align}
If we now consider the initial conditions \eqref{Init1}, we find 
$
(\partial \phi)^2 = j^{~\mu}_\nu\nabla_\mu\phi \nabla^\nu\phi \ ,
$
which is bounded on $\Omega$ by \eqref{BoundK}. This means that the term $\lambda h(\phi,(\partial \phi))$, which would appear in the energy $\mathcal{E}_\phi$, would still be bounded so it would not be able to counteract the $m^2$ term for large enough $m$ in $\mathcal{E}_\phi$. 
We omit detailed calculations as this case is a straightforward generalisation of the previous case but with longer expressions. In particular, the lower bound of the energy in the case $f(\phi,(\partial \phi)^2)$ would be different from the case $f(\phi)$ since $\mathcal{E}_{\mathrm{EMS}}$ in \eqref{EEMS} would include more terms from equation \eqref{wave-eq-2}.
%%%%%%%%%%%%%%%%%%%%%%%%%%%%%%%%%%%%%%%%%%
\section{The case for instability}
%%%%%%%%%%%%%%%%%%%%%%%%%%%%%%%%%%%%%%%%%%
\label{finite-energy}

Theorem \ref{main-thr} implies that the energy flux  either (i) diverges at $\mathscr{I}^+$ or (ii) is finite at $\mathscr{I}^+$. Then:

(i) The former case readily leads to an instability and assertion \eqref{instability} follows. 

(ii) In the latter case, arguments due to Friedman \cite{Friedman} and Moschidis \cite{Moschidis} indicate that this case is impossible. We will summarize their arguments next and leave the full details of the proof to a future publication.

In the Friedman-Moschidis approach, the impossibility of case (ii) is shown by contradiction as follows \cite{Friedman}:
Suppose $\phi$ is a solution to the wave equation having finite energy flux at $\mathscr{I}^+$. Then, it is shown in \cite{Moschidis} that $\phi$ "settles down" to a non-radiative state where the energy flux at $\mathscr{I}^+$ is zero, i.e.
\begin{equation}
\label{non-radiative}
\int_{\mathscr{I}^+} J_\mu^T(\phi, A^\alpha) n_{\mathscr{I}^+}^\mu=0.
\end{equation}
Therefore the unique continuation criterium of Assumption 1 ensures that $\phi\equiv 0$ on the whole spacetime. But this is in contradiction with assertion \eqref{negativity} of Theorem \ref{main-thr}, so the solution with finite energy flux at $\mathscr{I}^+$ is discarded. 

We now give some more details particularly regarding Moschidis' part of the proof which can be adapted to our case. Following the contradiction argument, the starting assumption is that all smooth solutions to the wave equation, on the domain of dependence of a Cauchy hypersurface $\Sigma_0$, with compactly supported initial data satisfy
\begin{equation}
\label{bound}
\sup_{u\ge 0} \int_{\Sigma_u} J_\mu^n(\phi, A^\alpha) n^\mu_{\Sigma_u}<+\infty.
\end{equation}
Then, there is the need to establish a convenient decay result for $\phi$ outside the ergoregion. In order to do that, Moschidis performs a frequency decomposition of the wave function $\phi$ into components with localised frequency support. Then, by providing weighted energy estimates for $\phi$, he established bounds in the frequency components \cite{Moschidis}. This part is based on the previous results \cite{Moschidis-2} which, crucially to our case, also apply to linear inhomogeneous wave equations of the type \eqref{wave-eq-2}. Then to show the decay and that $\phi$ actually vanishes outside the ergoregion, the essential ingredient are Carlman-type estimates for the solutions \cite{Moschidis}. In particular, given the resemblance to our problem, one should be able to show by similar methods that a sequence of solutions converges, weakly in $H^1_{loc}$ and strongly in $L^2_{loc}$, to the non-radiative state satisfying \eqref{non-radiative} thus providing the desired contradiction with \eqref{negativity}. But we haven't fully done that and leave further details to a future publication.
%%%%%%%%%%%%%%%%%%%%%%%%%%%%%%%%%%%%%%%%%%%%%%%%%%%%%%%%%

\section{Conclusion and future perspectives}
\label{conclusion}

We have investigated  the energy flux and the stability of stationary asymptotically flat spacetimes with an ergoregion and no future horizon in the context of Einstein-Maxwell-Scalar field models.
In order to do that we considered scalar field perturbations non-minimally coupled to a background electromagnetic field. Following ideas of J. Friedman \cite{Friedman} and G. Moschidis \cite{Moschidis}, we have shown that there is compactly supported initial data such that the energy flux across the initial Cauchy hypersurface is negative and that the energy flux decreases with time. This then indicates that the energy flux diverges and that those spacetimes are unstable.

An interesting question is whether the time scale of the instability would be small enough to be physically interesting or, on the contrary, it would be too large (e.g. larger than the age of the universe) and could be neglected. This may be investigated using numerical methods following the interesting results of \cite{Cardoso-2008, Shapiro-2019, Herdeiro-Santos, Cardoso-2022}.

On the other hand, the mathematical analysis of the non-linear system \eqref{wave-eq}-\eqref{wave-A} remains an open problem. The methods of Moschidis \cite{Moschidis} seem to be generalisable to wave equations with a mild non-linearity on a fixed background but a considerable amount of work would need to be done in particular for proving new decay estimates for the solutions.
 
Finally, as a third interesting line of research, we remark that one possible way to circumvent Theorem \ref{main-thr} is if the coupling function $f$ diverges. In that case, $h(\phi)$ will not be bounded and the instability can not be proved by the present strategy as the energy will be finite.
 In fact, solitonic solutions with a diverging $f(\phi)$ were found in \cite{Herdeiro:2019iwl,Herdeiro:2020iyi}. One such soliton solution was given for 
$
f(\phi)=1/(1-\alpha\phi)^4
$
with $\alpha$ constant, in which case although $f(\phi)$ diverges, the electromagnetic terms remain regular \cite{Herdeiro:2019iwl,Herdeiro:2020iyi}. One can thus conjecture that there may be cases where $f(\phi)$ is divergent and, nevertheless, one may have stable ergoregions in Einstein-Maxwell-Scalar models. 
%%%%%%%%%%%%%%%%%%%%%%%%%%%%%%%%%%%%%%%%%%%%%%%%%%%%%%%%%%%%%%%%%%
\section*{Acknowledgments}
%%%%%%%%%%%%%%%%%%%%%%%%%%%%%%%%%%%%%%%%%%%%%%%%%%%%%%%%%%%%%%%%%%
We thank G. Moschidis for very useful comments. We also thank support from CMAT, Univ. Minho, through FCT projects UIDB/00013/2020 and UIDP/00013/2020. FCM is supported by FCT/Portugal through CAMGSD, IST-ID, projects UIDB/04459/2020 and UIDP/04459/2020 and the H2020-MSCA-2022-SE project EinsteinWaves, GA No.101131233.
%%%%%%%%%%%%%%%%%%%%%%%%%%%%%%%%%%%%%%%%%%%%%%%%%%%%%%%%%%%%%%%%%%

%%%%%%%%%%%%%%%%%%%%%%%%%%%%%%%%%%%%%%%%%%%%%
\end{document}